\documentclass[aps,prl,english,twocolumn,showpacs,preprintnumbers,amsmath,amssymb,floatfix]{revtex4-1}

\usepackage[T1]{fontenc}
\usepackage[latin9]{inputenc}
\usepackage{graphicx}
\usepackage{amssymb}
\usepackage{babel}


\usepackage{babel}
\makeatother

\begin{document}

\title{Retardation turns the van der Waals attraction into a Casimir repulsion already at 3 nm}

\author{Mathias Bostr{\"o}m}
\author{Bo E. Sernelius}
\email{bos@ifm.liu.se}
\affiliation{Division of Theory and Modeling, Department of Physics, 
Chemistry
and Biology, Link\"{o}ping University, SE-581 83 Link\"{o}ping, Sweden}

\author{Iver Brevik}
\affiliation{Department of Energy and Process Engineering, Norwegian University of Science and Technology, N-7491 Trondheim, Norway}

\author{Barry W. Ninham}
\affiliation{Department of Applied Mathematics, Australian National University, Canberra, Australia}

\begin{abstract}
Casimir forces between surfaces immersed in bromobenzene have recently been measured by Munday et al. Attractive Casimir forces were found between gold surfaces. The forces were repulsive between gold and silica surfaces. We show the repulsion is due to retardation effects. The van der Waals interaction is attractive at all separations. The retardation driven repulsion sets in already at around 3 nm. To our knowledge retardation effects have never been found at such a small distance before. Retardation effects are usually associated with large distances.
\end{abstract}


\maketitle

When two objects are brought together correlations in  fluctuations of the charge and current densities in the objects or fluctuations of the fields usually result in an attractive force\,\cite{ Lond, Casi,Dzya,Lang, Maha, Isra, Ser, Milt, Pars, Ninhb}. At short distances this is the van der Waals force\,\cite{Lond}; at large distances the finite velocity of light becomes important (retardation effects) and the result is the Casimir force\,\cite{Casi, Milt}. With retardation effects we here mean all effects that appear because of the finite speed of light; not just the reduction in correlation between the charge density fluctuations at large distances; new propagating solutions to Maxwell's equations appear that bring correlations between current density fluctuations. When these propagating modes dominate we call the force Casimir force; when the surface modes\,\cite{Ser} dominate we call the force van der Waals force. The classical literature says that retardation is due to the finite
speed of light that weakens the correlations. This is misleading. It is due to the quantum nature of light\,\cite{Wenn,Ninha}; this becomes obvious at the large separation limit where the weakening is independent of the speed of light.

Since the famous experiments of Deryaguin and Abrikossova\,\cite{Der} there has been much interest in phenomena which measure the van der Waals forces acting between macroscopic bodies. The early experiments which measured the forces between quartz and metal plates covered only the retarded region. The experiments of Tabor and Winterton\,\cite{Tab} and subsequently of Israelachvili and Tabor\,\cite{IsraTabor,White} fitted the potential to a power law of $1/d^{n}$ ($d$ being the distance) where $n$ varied from non retarded ($n = 3$) to fully retarded ($n = 4$) value. There was a gradual transition from nonretarded  to retarded forces as the separation was increased from 12 nm to 130 nm\,\cite{IsraTabor}. A gradual transition in the case of two gold surfaces was demonstrated experimentally by Palasantzas et al.\,\cite{Palas} 

Casimir predicted already in 1948\,\cite{Casi} the attraction between a pair of parallel, closely spaced, perfect conductors a distance apart.
Almost 50 years later Lamoreaux\,\cite{Lamo} performed the first high accuracy measurement of Casimir forces between metal surfaces in vacuum. Following Tabor and Isrealachvili, and later Lamoreaux, a new research field developed\,\cite{Sush,Milt2}.

An interesting aspect of the Casimir force is that according to theory it can also be repulsive\,\cite{Dzya,Rich1,Rich}.   Early experiments showing this indirectly will be discussed at the end of this communication. Munday, Capasso, and Parsegian\,\cite{Mund} performed the first direct force measurements that demonstrated that the Casimir force could be repulsive by a suitable choice of interacting surfaces in a fluid. They found attractive Casimir forces between gold (Au) surfaces in bromobenzene (Bb). When one surface was replaced with silica (SiO$_2$) they measured a repulsive Casimir force. They speculated that this effect could allow quantum levitation of nano-scale devices\,\cite{Mund}. 

\begin{figure}
\includegraphics[width=7.55cm]{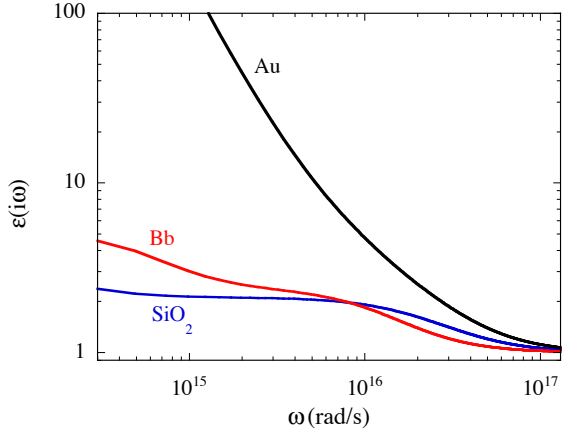}
\caption{(Color online) The dielectric function at imaginary frequencies for SiO$_2$ (silica)\,\cite{Grab}, Bb (bromobenzene)\,\cite{Mund}, and Au (gold)\,\cite{Bost2}.}
\label{figu1}
\end{figure}

We present calculations for the same system (SiO$_2$\,\cite{Grab} interacting with Au\,\cite{Bost2} across Bb\,\cite{Mund}) and find that attractive and repulsive Casimir forces are produced in one and the same system. Most notably we find that retardation effects changes the sign of the Casimir force already for separations of the order of 3-5 nm. Retardation effects usually come at much larger separations. 
The Casimir force can be calculated if the dielectric functions (for imaginary frequencies, $i\omega$) are known. These functions are shown in Fig.\,\ref{figu1}. 

There is a crossing between the curves for SiO$_2$ and Bb. This opens up for the possibility of a transition of the Casimir energy, from attraction to repulsion. 

The interaction of material 1 (SiO$_2$) with material 3 (Au) across medium 2 (Bb) results in a summation of imaginary frequency terms\,\cite{Ser}:
\begin{equation}
F = \sum\limits_{n = 0}^\infty  {{'}g\left( {{\omega _n}} \right)}.
\label{equ1}
\end{equation}
Note that positive values of $F$ correspond to repulsive interaction.

In the retarded treatment there are contributions from the two mode-types transverse magnetic (TM) and transverse electric (TE), $g\left( {{\omega _n}} \right) = {g^{TM}}\left( {{\omega _n}} \right) + {g^{TE}}\left( {{\omega _n}} \right)$, 
where
 \begin{equation}
\begin{array}{*{20}{l}}
{{g^{TM}}\left( {{\omega _n}} \right) = \frac{1}{\beta }\int {\frac{{{d^2}q}}{{{{\left( {2\pi } \right)}^2}}}ln\left\{ {1 - {e^{ - 2{\gamma _2}d}}\left[ {\frac{{{\varepsilon _1}\left( {i{\omega _n}} \right){\gamma _2} - {\varepsilon _2}\left( {i{\omega _n}} \right){\gamma _1}}}{{{\varepsilon _1}\left( {i{\omega _n}} \right){\gamma _2} + {\varepsilon _2}\left( {i{\omega _n}} \right){\gamma _1}}}} \right]} \right.} }\\
{\quad \quad \quad \quad \quad \quad \quad \quad \quad  \times \left. {\left[ {\frac{{{\varepsilon _3}\left( {i{\omega _n}} \right){\gamma _2} - {\varepsilon _2}\left( {i{\omega _n}} \right){\gamma _3}}}{{{\varepsilon _3}\left( {i{\omega _n}} \right){\gamma _2} + {\varepsilon _2}\left( {i{\omega _n}} \right){\gamma _3}}}} \right]} \right\};}\\
{{g^{TE}}\left( {{\omega _n}} \right) = \frac{1}{\beta }\int {\frac{{{d^2}q}}{{{{\left( {2\pi } \right)}^2}}}ln\left\{ {1 - {e^{ - 2{\gamma _2}d}}\left[ {\frac{{{\gamma _2} - {\gamma _1}}}{{{\gamma _2} + {\gamma _1}}}} \right]\left[ {\frac{{{\gamma _2} - {\gamma _3}}}{{{\gamma _2} + {\gamma _3}}}} \right]} \right\}} ;}\\
{{\gamma _i} = \sqrt {{q^2} - {\varepsilon _i}\left( {i{\omega _n}} \right){{\left( {i{\omega _n}/c} \right)}^2}} .}
\end{array}
\label{equ2}
\end{equation}
In the nonretarded treatments there are no TE contributions and the result is
\begin{equation}
\begin{array}{l}
g\left( {{\omega _n}} \right) = \frac{1}{\beta }\int {\frac{{{d^2}q}}{{{{\left( {2\pi } \right)}^2}}}ln\left\{ {1 - {e^{ - 2qd}}\left[ {\frac{{{\varepsilon _1}\left( {i{\omega _n}} \right) - {\varepsilon _2}\left( {i{\omega _n}} \right)}}{{{\varepsilon _1}\left( {i{\omega _n}} \right) + {\varepsilon _2}\left( {i{\omega _n}} \right)}}} \right]} \right.} \\
\left. {\quad \quad \quad \quad \quad \quad \quad \quad \times \left[ {\frac{{{\varepsilon _3}\left( {i{\omega _n}} \right) - {\varepsilon _2}\left( {i{\omega _n}} \right)}}{{{\varepsilon _3}\left( {i{\omega _n}} \right) + {\varepsilon _2}\left( {i{\omega _n}} \right)}}} \right]} \right\}.
\end{array}
\label{equ3}
\end{equation}

\begin{figure}
\includegraphics[width=7.55cm]{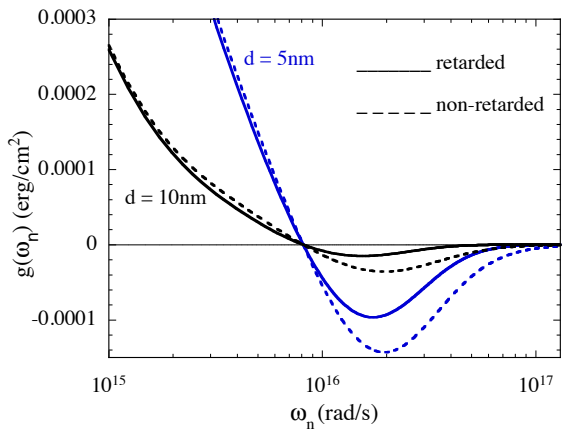}
\caption{(Color online)The spectral function,  $g\left( {{\omega _n}} \right)$, as a function of imaginary frequencies for two different separations.}
\label{figu2}
\end{figure}

\begin{figure}
\includegraphics[width=7.55cm]{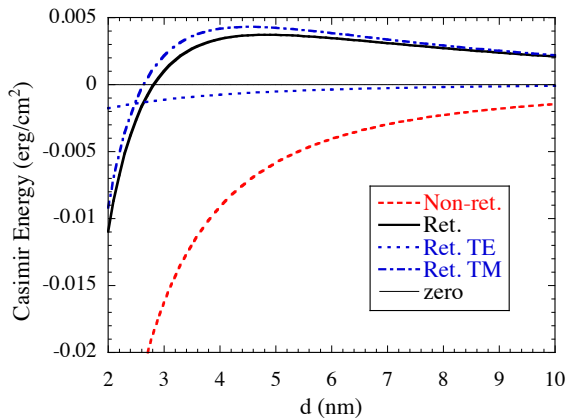}
\caption{(Color online) The retarded and nonretarded Casimir interaction free energy between gold and silica in bromobenzene. Also included are the contributions from TM and TE modes.}
\label{figu3}
\end{figure}

\begin{figure}
\includegraphics[width=7.55cm]{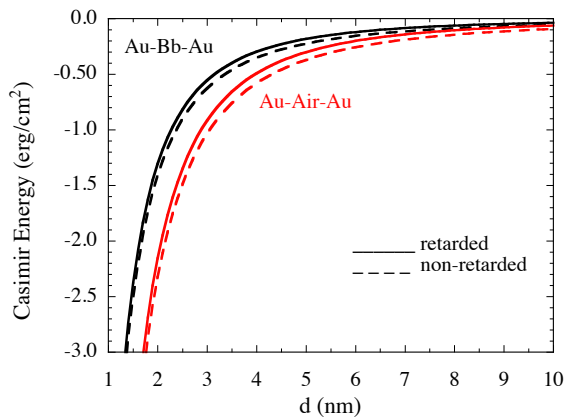}
\caption{(Color online) The retarded and nonretarded Casimir interaction free energy between two gold surfaces in either (a) bromobenzene or (b) air.}
\label{figu4}
\end{figure}

Frequency intervals where the intervening medium has a dielectric permittivity in between the permittivity of the two objects give a repulsive contribution; other intervals give an attractive contribution. This is obvious in the nonretarded treatment because of the factor $\left( {{\varepsilon _{Au}} - {\varepsilon _{Bb}}} \right)\left( {{\varepsilon _{Si{O_2}}} - {\varepsilon _{Bb}}} \right)$  appearing in the integrand. It is not so obvious in the retarded treatment where the integrands are more complex. The decomposition of the Casimir free energy into the frequency dependent and distance dependent spectral function, $g\left( {{\omega _n}} \right)$, is presented in Fig.\,\ref{figu2}. We note as for the systems above the reversal of the sign of the Casimir energy with retardation as compared to without retardation is due to a subtle balance of attractive (high frequencies) and repulsive (low frequencies) contributions. When the separation increases from its lowest value the attractive contributions at high frequencies are weakened by retardation at a higher rate than  the repulsive contributions at low frequencies.
The distance-dependent exponential,  $exp\left[ { - 2d\sqrt {{q^2} + {\varepsilon _{Bb}}\left( {i\omega } \right){{\left( {\omega /c} \right)}^2}} } \right]$, in the Lifshitz expression for the Casimir energy, Eqs.\,(\ref{equ2}) and (\ref{equ3}), has the effect that all frequencies contribute to the nonretarded van der Waals force while only low frequencies contribute in the long range retarded Casimir regime. The ultimate long-range asymptote, the entropic term, only includes the zero frequency transverse magnetic modes. Both the long-range Casimir asymptote and the long-range entropic asymptote are repulsive for the system considered here. However, as can be seen in Fig.\,\ref{figu3}. there is a transition region between the van der Waals region (where retardation does not influence the result) and the long range Casimir region. In this region the retarded Casimir energy is attractive. This means that it is possible to via the optical properties tune the materials used such that they in one region give attractive Casimir interactions and in another region give repulsive Casimir interactions. There is also a, to our knowledge not yet experimentally observed, maximum of the repulsive Casimir interaction. This maximum (which occurs around d = 5 nm) and the transition from attraction to repulsion (which occurs around d = 3 nm) pose suitable experimental tests for the validity of the theory. Notably the contributions from transverse magnetic modes are zero around approximately 26 \AA. This means that at one very short separation the Casimir force comes entirely from transverse electric modes. These modes are absent if retardation is neglected. This is a consequence of that the total force in the system is very small. Our results suggest that it is possible to measure retardation effects at much shorter separations than has previously been expected. Experiments are often performed in the sphere-plate geometry. Then the force in terms of the Casimir free energy, $F\left( d \right)$, is\,\cite{Der2} $2\pi R \cdot F\left( d \right)$, where we see that the force is proportional to the energy; the maximum repulsive force occurs at the position of the maximum repulsive energy.
The attractive Casimir energy between gold surfaces in either bromobenzene or air is shown in Fig.\,\ref{figu4}.   
Here both the retarded and nonretarded energies decrease monotonically with separation. 

\begin{figure}
\includegraphics[width=7.55cm]{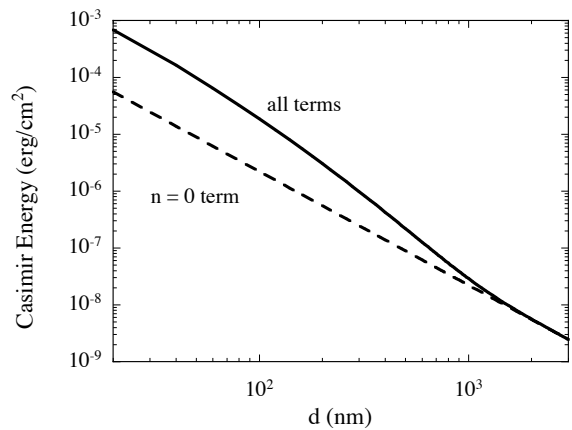}
\caption{The fully retarded Casimir interaction free energy and the n = 0 entropic term for a system with Au interacting with SiO$_2$ across bromobenzene.}
\label{figu5}
\end{figure}

The long-range part of the interaction is shown in Fig.\,\ref{figu5}.  
In the 1-2 $\mu$m range the thermal entropic effect starts to become important. This is similar to what we found for the Casimir force between real metal surfaces in vacuum\,\cite{Bost2}. 	

Using optical measurements on the interacting objects and liquid makes it possible to predict the force --- from short-range attractive Casimir force to long range repulsive Casimir force. We find similar effects of retardation when gold is replaced with silver (the difference is that the crossing from attractive to repulsive Casimir force then comes around d = 5 nm). 
We predict strong retardation effects that are much more pronounced at short and intermediate separations in systems with repulsive Casimir interaction as compared to in systems with attractive Casimir interaction. We stress that it should be possible to measure a maximum repulsive Casimir force. We suggest that this distance of maximal repulsion should be the optimal distance to enable quantum levitation of nano-scale devices (floating on each other in a liquid due to repulsive Casimir interactions)\,\cite{Mund}. Another system that changes sign for a particular separation was discussed by Munday et al.\,\cite{Mund2} for birefringent disks in a fluid; however, there was little discussion regarding the significance of transition.

We would finally like to mention  a caveat that poses inevitable  difficulties in interpretation of measurement of  these forces.  With isotropic molecules the granularity of liquids near surfaces gives rise to oscillatory forces--much larger than van der Waals--extending to at least 4-6 molecular layers before merging into continuum theory\,\cite{Ninhb,Isra,Qua,Bona}. With anisotropic molecules, overlap of surface induced order parameters gives rise to attractive or repulsive forces of the same range. If they do not turn up, the reasons may be that the surfaces used are rough at a molecular level. This tends to smooth oscillations. The minimum roughness would be say around 2 molecular layers at each surface. This gives rise to an uncertainty in distance  that must be considered in calculations that aim at comparing with high accuracy experiments.

To sum up, we have seen that the effects of retardation turn up already at very short distances, of the order of a few nm. Remarkably the effect of retardation can be to turn attraction into repulsion. The essential physical parameter is $\varepsilon(i\omega_n)$, a real and positive quantity intimately connected with absorption (imaginary permittivity $\varepsilon''$) through the Kramers-Kronig dispersion relation
\begin{equation}
\varepsilon(i\omega_n)-1=\frac{2}{\pi}\int_0^\infty \frac{x\varepsilon''(x)}{x^2+\omega_n^2}\,dx.
\end{equation}
From this perspective it is thus the coupling between the system to a frequency-dependent heat bath, rather than the finite speed of light, which is important. Repulsive Lifshitz forces were found a long time ago in indirect force measurements. One example is  the work of Anderson and Sabiski on films of liquid helium on calcium fluorite, and similar molecularly smooth surfaces \,\cite{AndSab}. The films ranged from 10-200 \AA \,\cite{AndSab}. The thickness of the films could be measured to within a few percent in most cases. For the saturated-film measurements the repulsive van der Waals potential was equal to the negative of the gravitational potential \,\cite{AndSab}. A good agreement was found\,\cite{Rich1} between these experimental data and the results from Lifshitz theory. In another more subtle experiment Hauxwell and Ottewill\,\cite{Haux} measured the thickness of films of oil on water near the alkane saturated vapour pressure. For this system n-alkanes up to octane spread on water. Higher alkanes do not spread. It is an asymmetric system (oil-water-air) and the surfaces are molecularly smooth. The phenomenon depends on a balance of van der Waals forces against the vapor pressure\,\cite{Rich, Haux, Ninh}. The net force, as a function of film thickness depends on the dielectric properties of the oils. As demonstrated\,\cite{Rich} it involves an intricate balance of repulsive and attractive components from different frequency regimes. When the ultraviolet and visible components are exponentially damped by retardation the opposing (repulsive) infrared components take over\,\cite{Ninh}.
  
 It is instructive finally to compare this with the theory for surface melting of ice \cite{Elba}. Surface melting takes place if the liquid phase wets the solid-vapor interface at the triple point. Again, it is the delicate balance between the permittivities $\varepsilon(i\omega_n)$ for the liquid and the solid which is the main factor. While water is more polarizable than ice at optical frequencies, the permittivity for ice exceeds that of water for frequencies higher than about $2\times 10^{16}$ rad/s. A thin layer of water at first tends to thicken, but the thickening becomes gradually hampered because of the retardation. The surface melting is thus incomplete. One notices from Fig.~1 in \cite{Elba} that the crossing point between the permittivities for ice and water occurs at about $1.5\times 10^{16}$ rad/s, thus about at the same point as the crossing point in Fig.~1 above.

\begin{acknowledgments}
The research was sponsored by the VR-contract No:70529001 which is gratefully acknowledged.
\end{acknowledgments}

\end{document}